\documentclass[aps,prb,twocolumn,showpacs]{revtex4}
\usepackage[latin1]{inputenc}
\usepackage{eso-pic}
\usepackage{graphicx}
\usepackage{color}
\usepackage{amsmath}
\usepackage{amssymb}
\usepackage{xcolor}
\usepackage{soul}

\usepackage{hyperref}
\hypersetup{
    colorlinks=true, 
    linktoc=all,     
    citecolor=black,
    filecolor=black,
    linkcolor=black,
    urlcolor=black
}

\begin{document}

\title{Emission Redshift in DCM2-Doped Alq$_{3}$ Caused by Non-Linear Stark Shifts and F\"{o}rster-Mediated Exciton Diffusion }

\author{Grazi\^{a}ni Candiotto\footnote{G.C. and R.G. contributed equally for this work.}\footnote{gcandiotto@iq.ufrj.br }$^{1,2}$}
\author{Ronaldo Giro$^{*}$\footnote{rgiro@br.ibm.com}$^{3}$}
\author{Bruno A. C. Horta$^{2}$}
\author{Fl\'{a}via P. Rosselli$^{4,6}$}
\author{Marcelo de Cicco$^{4}$}
\author{Carlos A. Achete$^{4}$}
\author{Marco Cremona$^{4,5}$ }
\author{Rodrigo B. Capaz\footnote{capaz@if.ufrj.br}$^{1,4}$}
\affiliation{$^{1}$ Instituto de F\'{i}sica, Universidade Federal do Rio de Janeiro, Rio de Janeiro, RJ 21941-972,  Brazil}
\affiliation{$^{2}$ Institute de Qu\'{i}mica, Universidade Federal do Rio de Janeiro, Rio de Janeiro, 21941-909, Brazil}
\affiliation{$^{3}$ IBM Research, Av. Pasteur 138/146, 22290-240, Rio de Janeiro-RJ, Brazil}
\affiliation{$^{4}$ Instituto Nacional de Metrologia, Qualidade e Tecnologia (INMETRO), 25250-020, Duque de Caxias, Brazil}
\affiliation{$^{5}$ LOEM, Departamento de F\'{i}sica, Pontif\'{i}cia Universidade Cat\'{o}lica do Rio de Janeiro, 22451-900, Rio de Janeiro, Brazil}
\affiliation{$^{6}$ Departamento de Qu\'{i}mica, Universidade Federal de S\~{a}o Carlos, S\~{a}o Carlos-SP, 13565-905, Brazil}

\date{\today}

\begin{abstract}
Organic light-emitting diodes (OLEDs) devices in the archetype small molecule fluorescent guest-host system tris(8-hydroxyquinolinato) aluminum (Alq$_{3}$) doped with 4-(dicyanomethylene)-2-methyl-6-julolidyl-9-enyl-4H-pyran (DCM2) displays a redshift in light-emission frequency which is extremely sensitive to the dopant concentration. This effect can be used to tune the emission frequency in this particular class of OLEDs. In this work, a model is proposed to describe this effect using a combination of density functional theory (DFT) quantum-chemical calculations and stochastic simulations of exciton diffusion via a F\"orster mechanism. The results show that the permanent dipole moments of the Alq$_{3}$ molecules generate random electric fields that are large enough to cause a non-linear Stark shift in the band gap of neighboring DCM2 molecules. As a consequence of these non-linear shifts, a non-Gaussian probability distribution of highest-occupied molecular orbital to lowest-unoccupied molecular orbital (HOMO-LUMO) gaps for the DCM2 molecules in the Alq$_{3}$ matrix is observed, with long exponential tails to the low-energy side. Surprisingly, this probability distribution of DCM2 HOMO-LUMO gaps is virtually independent of DCM2 concentration into Alq$_{3}$ matrix, at least up to a fraction of 10\%. This study shows that this distribution of gaps, combined with out-of-equilibrium exciton diffusion among DCM2 molecules, are sufficient to explain the experimentally observed emission redshift. 
\end{abstract}


\maketitle

\section{Introduction}

Organic light emitting diodes (OLEDs) are a relatively new class of devices already used for display technologies \cite{brad+93sm,sheats+96} (TV, computer, cell phones, palmtop computer screens etc.)\citep{ma2019,nuesch2005,soman2020} and other applications as illumination source\cite{kamte+10am}, lasers\cite{bulovic1998weak,vcehovski2020,zhang2011existence} and medical devices \cite{chin2005medical}. The great interest in this technology is related to the low cost of organic materials, simplicity of organic thin film growth, ease of integrability with conventional technology, versatility of carbon chemistry, among other advantages. However, this technology has some drawbacks, as the device lifetime still needs to be improved and OLEDs show generally a broad electroluminescent (EL) resulting in unsaturated emission colors.

\begin{figure}[htbp]
\centerline{\includegraphics[width=9.0 cm]{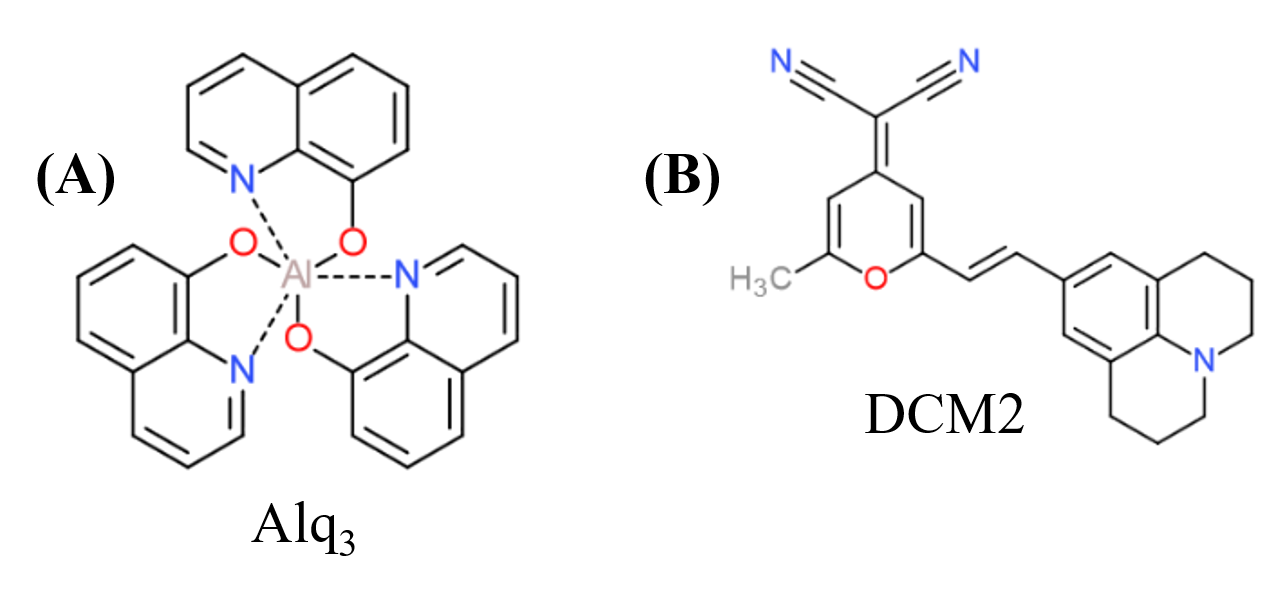}}
\caption{Schematic representation of (A) Alq$_{3}$, and (B) DCM2 molecules.} 
\label{dcm2-alq3}
\end{figure}

In order to overcome the latter problem, Bulovi\'{c} \textit{et al}.\cite{bulovic+98cpl,madi+03prl} developed OLED devices by doping a "host" material aluminum tris(8-hydroxy quinoline) (Alq$_{3}$) (Fig. \ref{dcm2-alq3} (A)) with "guest" molecules [2-methyl-6-[2-(2,3,6,7-tetrahydro-1H, 5H-benzo
[i,j] quinolizin-9-yl)-ethenyl]-4H-pyran-4-ylidene] propane-dinitrile (DCM2) (Fig. \ref{dcm2-alq3}(B)), hereafter referred to as Alq$_{3}$:DCM2. In these devices, excitons are generated in the Alq$_{3}$ molecules and efficiently transferred by F\"{o}rster resonance energy transfer (FRET)\cite{forster+48ap,medi+2014,vcehovski2020} to the DCM2 molecules. Moreover, these devices show saturated color emission\cite{cheon2004electroluminescence}, and the color can vary from yellow to red as the concentration of guest (DCM2) molecules is increased from 1\% to 10\%. This redshift amounts to roughly 50 nm, with a relatively unchanged peak width over this range of doping\cite{bulovic+98cpl,madi+03prl,bulovic1998weak}. Due to its interesting properties and numerous applications in the area of organic thin films, the Alq$_{3}$:DCM2 system is still a relevant topic which draws the attention of the scientific community\cite{vcehovski2020,liu2014temperature,becker2017}.

Previous works suggested that the spectral shift was due to excimer formation \cite{tang+89jap,madi+03prl} or hydrogen bonds in solution\cite{kali+94cp}. Bulovi\'{c} \textit{et al}. \cite{bulovic+98cpl, madi+03prl} challenged these interpretations, because excimer formation would not result in a rigid and continuous shift of the electroluminescence (EL) spectrum. In addition, hydrogen bonds with DCM2 molecules are not possible in solution\cite{bulovic+98cpl, madi+03prl}. The similarity of the spectral widths, and magnitudes of the peak shifts of the spectra both in solution and in thin films suggested Bulovi\'{c} \textit{et al}.\cite{bulovic+98cpl,madi+03prl} to attribute the redshift in energy due to the polarization induced by DCM2 molecules. In their own words: \textit{"as the DCM2 concentration in the relatively non-polar Alq$_{3}$ is increased, the distance between nearest neighbor, highly polar, DCM2 molecules decreases, thereby increasing the local polarization field. This polarization tends to redshift the DCM2 emission spectrum."} \cite{bulovic+98cpl,madi+03prl}

In a series of articles\cite{bulovic+98cpl, bulovic+99cpl, madi+03prl}, Bulovi\'{c} and collaborators called this solid-state solvation effect (SSSE), in analogy to the "solvation effect" of organic dyes in liquid solutions, which is observed when the dye absorption and emission spectra are influenced by dipole moment of the surrounding solvent molecules\cite{reichardt2011solvents}. The "solvation effect" results from inter-molecular solute-solvent interaction forces such as dipole-dipole or dipole-induced dipole (these interactions tend to alter the energy difference between the ground and excited state of the solute). The SSSE effect has been used for tuning the luminescent emission spectrum of dipolar molecules by adjusting the strength of intermolecular dipole-dipole interactions using a doped guest-host molecular organic thin film system\cite{bulovic+99cpl, madi+03prl}.

Bulovi\'{c} \textit{et al}.\cite{bulovic+98cpl, madi+03prl} also made an important observation: Since the molecules in the solid solution must be randomly distributed, over a large volume, the net DCM2 dipole moment averages to zero. However, considering that the dipole field decreases as $1/r^{3}$ , where $r$ is the distance between dipoles, near any given radiating molecule there should be a net local electric field due to the dipole moments of neighboring DCM2 molecules which, on average, influences the spectral emission.

Other models have been suggested\cite{baldo+01cpl,madi+03prl} to explain the observed redshift.
In 2001, Baldo \textit{et al}. \cite{baldo+01cpl} introduced the so-called "local order theory". This theory is based on the formation of aggregates of guest molecules into the host matrix. Baldo \textit{et al}. argued that as the DCM2 concentration increases from 1\% to 10\%, the DCM2 molecules readily aggregate. The spectral shifts are then explained due to the high electric fields associated with local ordering of the polar DCM2 molecules in aggregate domains. In a following work, Madigan \textit{et al}. \cite{madi+03prl} developed a model of solvatochromism relating the experimentally observed changes in emission and absorption spectra of a solute to the electronic permittivity of a solvent. This model does not require the assumption of aggregation of DCM2 to explain the redshift and it was supported by experimental data \cite{madi+03prl}.

Regardless of whether the spectral redshift is related to aggregation or not, all previous models relied on the fact that emission spectra would vary with changes of local electric field due to high electric dipole moment and dielectric constant of DCM2, as its concentration increases from 1\% to 10\%. However, the detailed mechanism for this effect was not investigated at the level of quantum-chemical calculations. In particular, it is puzzling the association of a strong emission redshift to an electric field acting on the DMC2 molecules, since changes in the electronic or optical gap under an electric field (Stark shifts) are typically linear to first order. Therefore a randomly-oriented field should in principle give rise to both positive and negative variations of the gap, with a nearly zero net effect.

In the present work, we develop a new model to explain the redshift emission in DCM2-doped Alq$_{3}$, supported by a combination of DFT quantum-chemical calculations and stochastic simulations of exciton diffusion and emission. Surprisingly, the energy gap distribution of DCM2 molecules under a random distribution of DCM2 and Alq$_{3}$ dipoles is rather independent of the DCM2 concentrations (for up to 10\% DCM2), as just the smaller dipole moments of neighboring Alq$_{3}$ molecules are sufficient to produce the necessary gap variations in DCM2 to account for the observed redshift. Moreover, the calculated Stark shifts are highly nonlinear, producing a probability distribution of DCM2 gaps with a long tail to the low-energy side. Finally, the observed concentration dependence of redshift is explained by exciton diffusion via the FRET mechanism. 

The paper is organized as follows: Section \ref{stark} describes quantum-chemical calculations of the HOMO-LUMO gap variations of DCM2 under electric fields (Stark shift). Section \ref{efdist}  presents simulations of the local electric field on DCM2 and the determination of the gap distribution using random Alq$_{3}$ and DCM2 dipoles distribution. Section \ref{kmc}, describes the kinetic Monte-Carlo (kMC) simulations of exciton diffusion and emission via F\"{o}ster energy transfer. Finally, Section \ref{conc} presents the main conclusions of the present work.

\section{Stark shift}
\label{stark}
\subsection{Methodology}

Since the random distribution of dipole moments of Alq$_{3}$ and DCM2 molecules results in an effective electric field acting on the DCM2 dopant molecules, initially we established the dependence of the DCM2 HOMO-LUMO gap as a function of the intensity and orientation of this field, \textit{i.e.}, the Stark shift is evaluated. To establish this dependence and at the same time ensure that our approach has quantitative and predictive capabilities, the molecular geometry, dipole moment, polarizability tensor, and HOMO-LUMO gap must be calculated using \textit{ab initio} methods. The quantum-chemical calculations were performed using the Gaussian03 program \cite{gaussian}. For the optimization of the geometry, dipole moment and polarizability tensor of DCM2 and Alq$_{3}$ molecules, the hybrid functional PBE1PBE\citep{perdew+96prl,perdew+97prl,adamo+99jcp} was used, along with the 6-31G(d,p)\cite{ditch+71jcp} basis set.

Once the geometry of the DCM2 molecular structure was optimized (see Fig. \ref{stark-effect}(A)), electric fields $\vec{E}$ of various intensities were applied in different orientations with respect to DCM2 molecules (see Fig. \ref{stark-effect}(B)). For these calculations, six different directions were chosen ($E_{x}$, $E_{y}$, $E_{z}$, $E_{xy}$, $E_{yz}$ and $E_{xz}$ respectively): Parallel to $x$, $y$ and $z$ axes, and at 45$^{\circ}$ with respect to $x$-axis in the $xy$ plane, at 45$^{\circ}$ with respect to $y$-axis in the $yz$ plane, and finally at 45$^{\circ}$ with respect to $z$-axis in the $xz$ plane. For this study, we performed SCF calculations using Gaussian03\cite{gaussian} within DFT. In this case the B3LYP\cite{B3,LYP,b3lyp-1,b3lyp-2} hybrid functional was used for the exchange-correlation term in DFT, with the same base set as in geometry optimization. The self-consistent field (SCF) calculations for this case is justified because it is expected that, in solid state film, DCM2 molecules in the Alq$_{3}$ matrix do not have enough space to accommodate geometry relaxation.

\subsection{Results}

\begin{figure}[htbp]
\centerline{\includegraphics[width=9.0 cm]{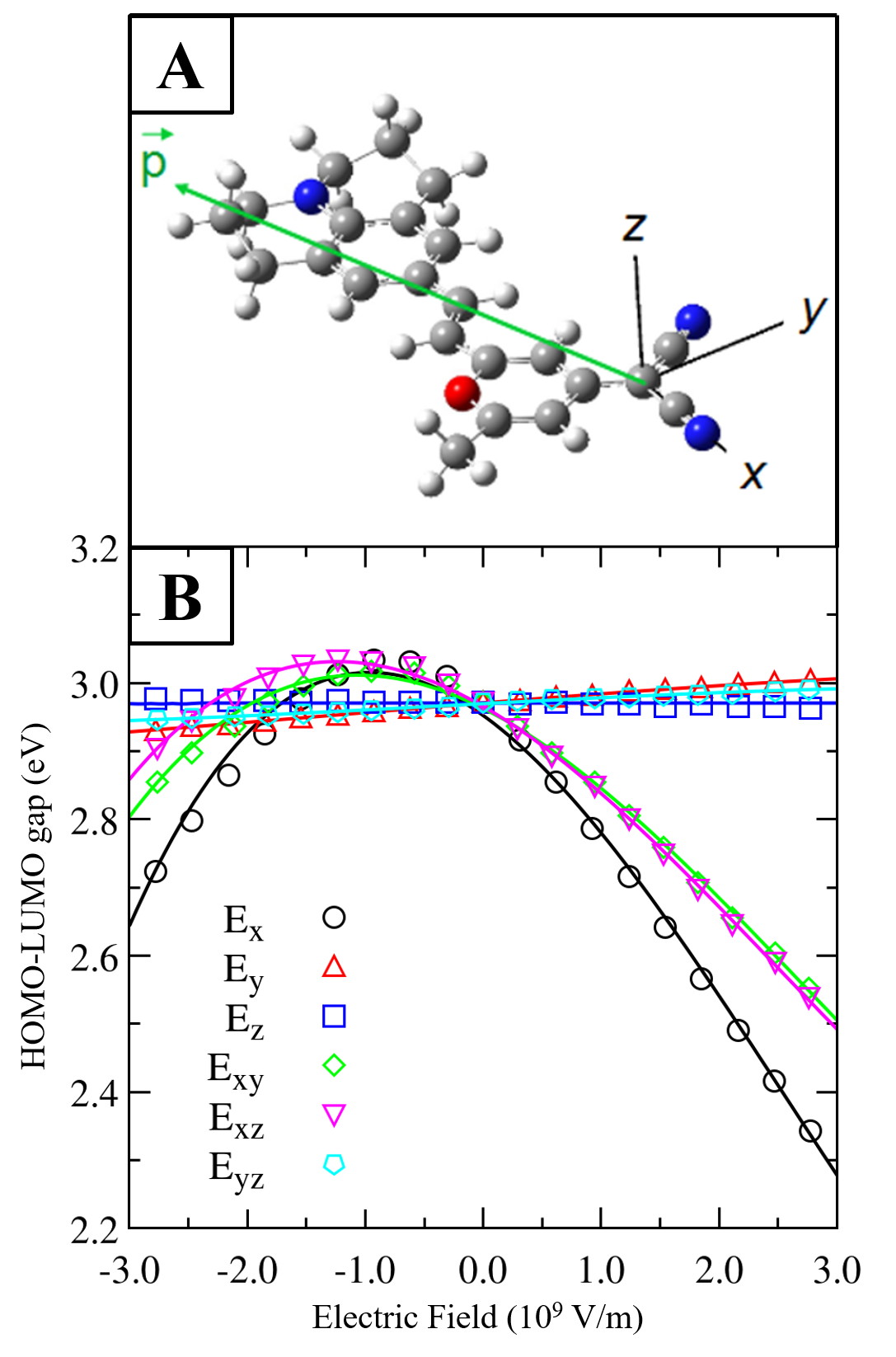}}
\caption{(A) Optimized geometry of DCM2 molecule. Almost all atoms of DCM2 molecule are lying in the $xy$ plane. Grey, white, blue and red spheres represent carbon, hydrogen, nitrogen and oxygen atoms, respectively. The arrow shows the orientation of dipole moment. The calculated DCM2 dipole moment vector is $\vec{p}_{p}=-12.6 \hat{x} -6.9 \hat{y} -0.2 \hat{z}$ D (Debye units). (B) DCM2 HOMO-LUMO gap as a function of applied electric field along directions $x$, $y$ and $z$. Solid lines represent the respective fit for each direction of the electrical field. Notice that the $x$ component shows a stronger and nonlinear dependence in the gap. The $x$ direction is almost parallel to the DCM2 dipole moment}.
\label{stark-effect}
\end{figure}

Based on the \textit{ab initio} DFT approach described previously, we first calculate the dipole moment and the polarizability tensor of Alq$_{3}$ and DCM2 molecules. These properties will be used in Section \ref{efdist}. As expected, DCM2 molecules are highly polar, with a ground state dipole moment of 14.4 D, as compared to the Alq$_{3}$ dipole moment of 4.4 D. These values are in good agreement with those reported in the literature \cite{bulovic+99cpl,curioni+98cpl}.  In Fig. \ref{stark-effect}(A) we show the optimized DCM2 geometry and the dipole moment vector. As one can see, the dipole moment is oriented from the two carbon-nitrogen groups towards the oxygen atom. This is due to the the balance of electronic charge between oxygen (negative) and the two carbon-nitrogen groups (positive) at the ends of the DCM2 molecule. 

Fig. \ref{stark-effect} (B) shows the dependence of DCM2 HOMO-LUMO gap as a function of the electric field (Stark effect). The Stark effect is stronger when the electric field is applied parallel to the $x$ direction ($E_{x}$ - electric field in $x$ direction). In this case, this effect has a nonlinear dependence. We also show the dependence of DCM2 gap regarding to variations of the applied electric field in other directions. For $E_{y}$, $E_{z}$ and $E_{yz}$ directions there is almost no variation of DCM2 gaps with respect to the electric field, but for $E_{xy}$ and $E_{xz}$ directions a similar behavior is observed as in $E_{x}$ direction.

Due to the nonlinear behavior shown in Fig. \ref{stark-effect}(B), an analytical expression for the HOMO-LUMO gap of DMC2 $E_{g}$ as function of the electric field needs to consider up to the quadratic terms:
\begin{eqnarray}\nonumber
E_{g}(\vec{E})= E_{g}(0)+\vec{\alpha}.\vec{E}+\vec{E}^t.\boldsymbol{\beta}.\vec{E}= \\ \nonumber E_{g}(0)+\alpha_{x}E_{x}+\alpha_{y}E_{y}+\alpha_{z}E_{z}+\beta_{xx}E_{x}^2+\beta_{yy}E_{y}^2 \\ 
+\beta_{zz}E_{z}^2+2\beta_{xy}E_{x}E_{y}+2\beta_{xz}E_{x}E_{z}+2\beta_{yz}E_{y}E_{z}
\label{eq-gap}
\end{eqnarray}

\noindent
where $E_{g}(0)$ = 2.96 eV is the HOMO-LUMO gap for the ground state at zero electric field. The coefficients $\alpha_i$ and $\beta_{ij}$ are obtained by fitting the DCM2 HOMO-LUMO gap dependence for each direction of the applied electric field shown in Fig. \ref{stark-effect} (B) by quadratic polynomials. The resulting the values of $\alpha_i$ and $\beta_{ij}$ are shown in Table \ref{tab-egfunction}. 

\begin{table}[h!]
\centering
\caption{Values of $\alpha_{i}$ and $\beta_{ij}$ terms from Eq. \ref{eq-gap}. These values are obtained from the coefficients of polynomial fitting from DCM2 gap dependence of applied electric field (see Fig. \ref{stark-effect}(B)).}
\begin{tabular}{ccc}
\hline
\hline
Direction             &    $\alpha_{i} $                 &  $ \beta_{ij}$     \\
$\hat{i}$             &              (eV.(V/m))                    &  (eV.(V/m))$^{2}$  \\
\hline
$\hat{x}$             &   $          -1.24800\times 10^{-10}$    &  $         -5.46220\times 10^{-20}$                 \\
$\hat{y}$             &   $ \,\,\,\,  1.32409\times 10^{-11}$    &  $         -3.89083\times 10^{-22}$                  \\
$\hat{z}$             &  $          -1.76392\times 10^{-12}$    &  $         -7.73069\times 10^{-23}$                  \\
$\hat{x}+\hat{y}$    &                                          &  $         -3.42360\times 10^{-20}$                \\
$\hat{x}+\hat{z}$     &                                          &  $         -3.22504\times 10^{-20}$                \\
$\hat{y}+\hat{z}$     &                                           &  $         -1.59986\times 10^{-22}$                 \\
\hline
\hline
\end{tabular}
\label{tab-egfunction}
\end{table}

\section{ELECTRIC FIELD AND ENERGY GAP DISTRIBUTIONS}
\label{efdist}

In this Section, the resulting electric field at each DCM2 molecule caused by a random distribution of Alq$_{3}$ and DCM2 dipole moments is calculated. Once obtained this electric field, the DCM2 gap shift is calculated using Eq. \ref{eq-gap}. With this procedure it is possible to obtain the histogram of DCM2 gap distribution for each concentration of DCM2 molecules into Alq$_{3}$ matrix.

To calculate the resulting electric field in each of DCM2 molecules, we consider not only the permanent dipole moments of Alq$_{3}$ and DCM2 molecules, but also the induced dipole moment due to polarization. The electric field calculation then follows a self-consistent iterative procedure, as illustrated in Fig. \ref{flow-chart} (more details can be found in the Supplemental Material).
\begin{figure}[htbp]
\centerline{\includegraphics[width=8.5 cm]{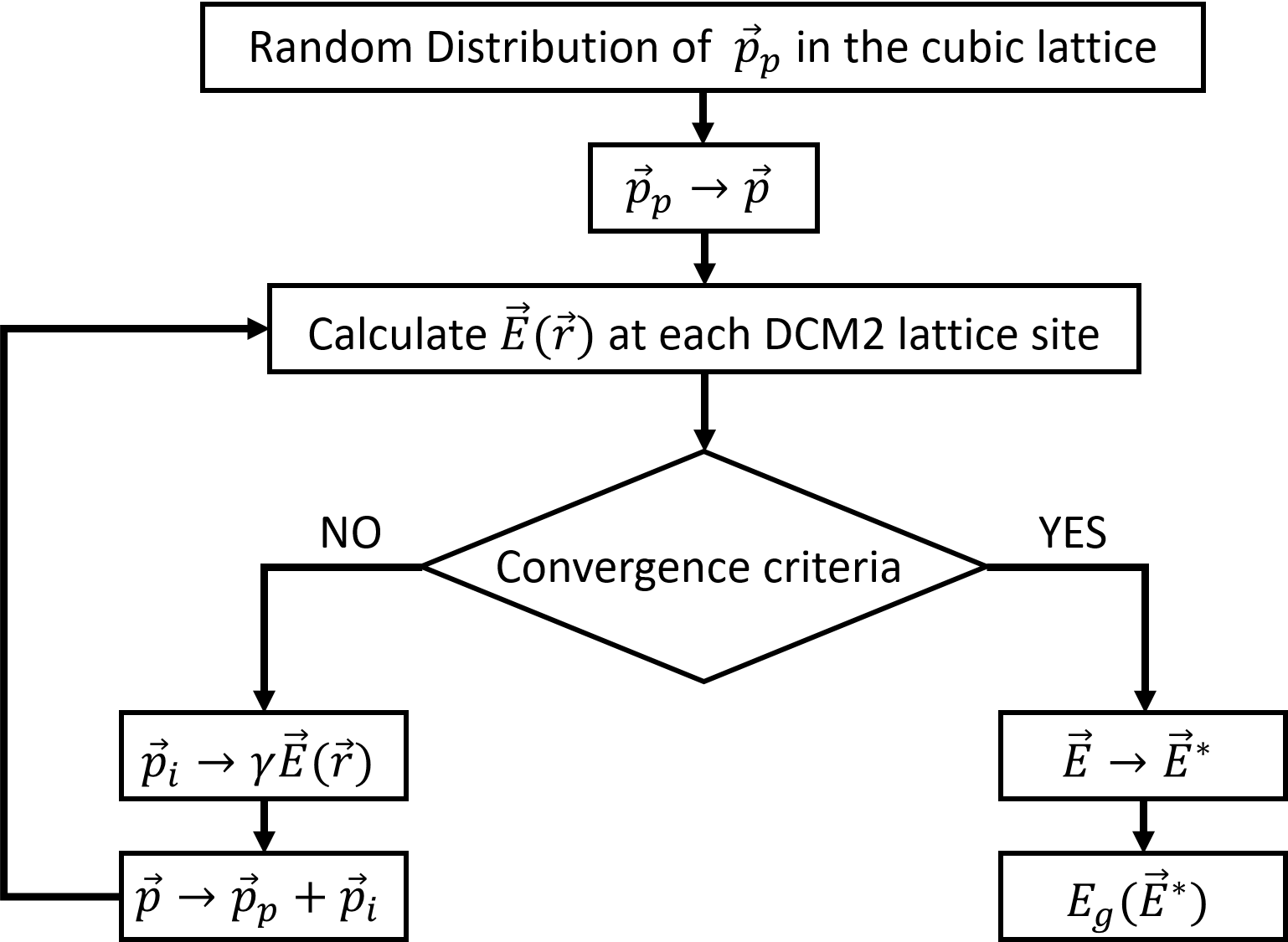}}
\caption{Flowchart illustrating the self-consistent procedure used in this work to calculate the DCM2 energy gap distribution due to the random electric field generated by the Alq$_3$ and DCM2 permanent and induced dipoles. $\vec{p}$, $\vec{p_p}$ and $\vec{p_i}$ indicate the total, permanent and induced dipole moments of each molecule, $\gamma$ is the polarizability tensor, $\vec{E}$ is the local electric field and $\vec{E^*}$ represents its converged value.} 
\label{flow-chart}
\end{figure}

In this methodology, Alq$_{3}$ and  DCM2  permanent  dipole  moments initially are distributed in a 70$\times$70$\times$70 cubic lattice, with lattice constant of 8.5 \AA. This lattice constant is chosen in order to reproduce the same density as amorphous Alq$_{3}$ matrix. The ratio of DCM2 and Alq$_{3}$ dipoles is selected respecting the DCM2 concentration in the Alq$_{3}$ host. All dipole moments are randomly oriented. In the second step, we calculated the electric field at each DCM2 and Alq$_{3}$ molecules due to the random distribution of dipoles. Thus the  induced dipole  moment  on  each  molecule  is  obtained from the calculated polarizability tensor and the total dipole moment is obtained as the sum of induced and permanent moments. Then the electric fields are recalculated and the convergence criteria are analyzed. The iterative process repeats until convergence is achieved. After convergence, the DCM2 gaps are calculated using Eq. \ref{eq-gap}.

\begin{figure}[htbp]
\centerline{\includegraphics[width=9.0 cm]{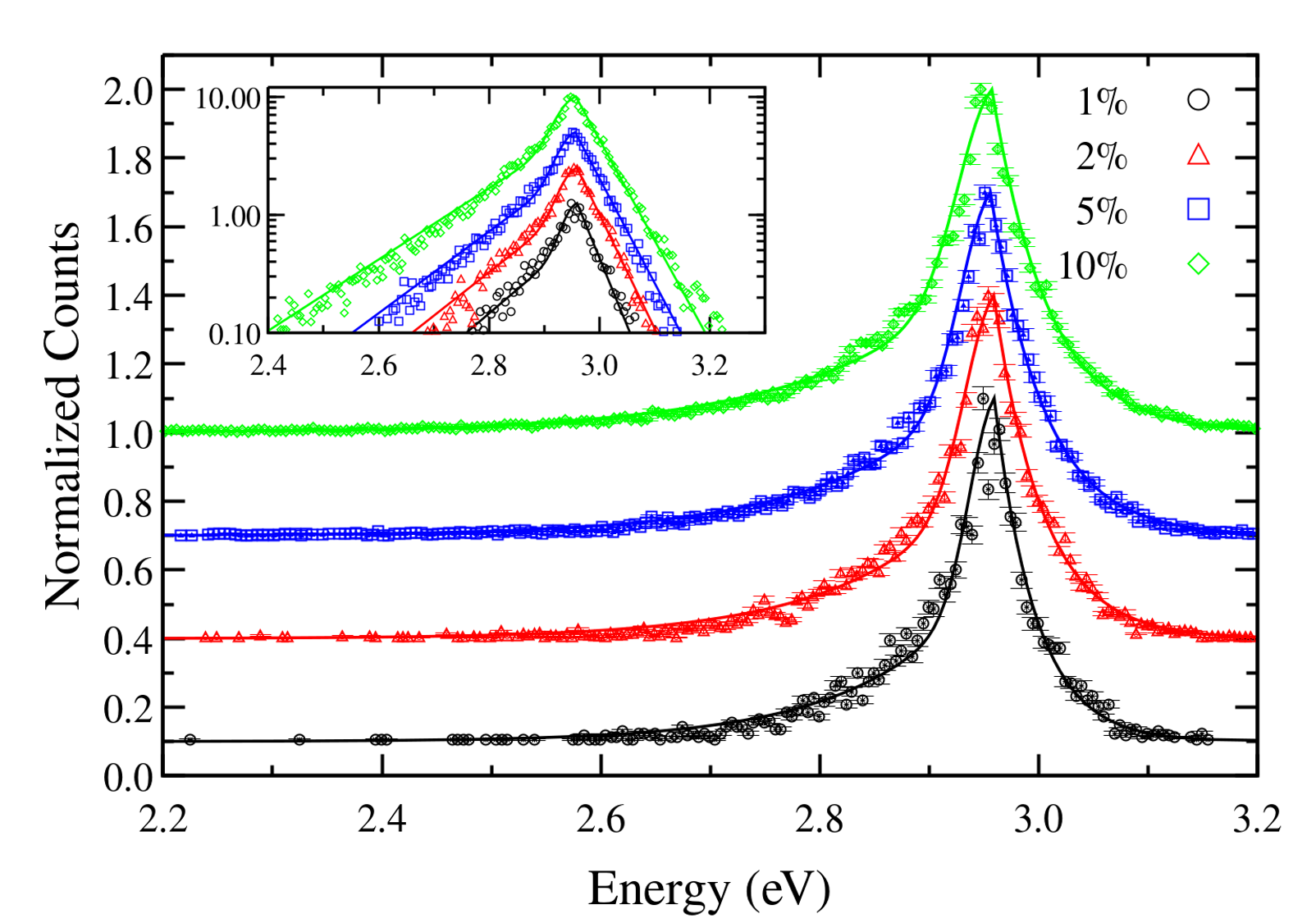}}
\caption{HOMO-LUMO gap distribution for different DCM2 concentrations. Points represent the histogram of DCM2 gap, and lines the fitted gap distribution function. In the detail, the HOMO-LUMO gap distribution are represented in logarithmic scale. Differents curves represent various DCM2 concentration in Alq$_{3}$ matrix and they are displaced vertically for clarity. Solid lines represent the respective fit for each DCM2 concentration}.
\label{fig-gap-distribution}
\end{figure}

The result of this procedure is shown in Fig. \ref{fig-gap-distribution} as a histogram showing the probability distribution of DCM2 HOMO-LUMO gaps. The DCM2 gap distribution is asymmetric, with a long tail in the low energy region. This is a direct consequence of the nonlinearity of the Stark shifts (see Fig. \ref{stark-effect}(A)). For energies lower than $E_{0}$=2.96 eV, the gap distribution shows a behavior that is approximately a linear combination of a Gaussian and an exponential function. For energies higher than $E_{0}$ the behavior is approximately exponential. Based on these empirical behaviors, it is possible to write an analytical expression for the probability distribution of the DCM2 gap (to be used in Section \ref{kmc}). The expression for the probability distribution is: 

{\footnotesize
\begin{equation}
P(E_g) = \begin{cases}
A\exp[\frac{-(E_{g}-E_{0})^2}{2\sigma^2}] + B\exp[\frac{(E_{g}-E_{0})}{\epsilon_{1}}] &\text{if } E_{g} \leq E_{0}\\
C\exp[\frac{-(E_{g}-E_{0})}{\epsilon_{2}}] &\text{if } E_{g} > E_{0} 
\end{cases}
\label{eq-probdcm2gap}
\end{equation}
}

\noindent where $A$, $B$ and $C$ are normalization constants, $E_g$ is the DCM2 energy gap distribution and $E_0$, $\sigma$, $\epsilon_{1}$ and $\epsilon_{2}$ are free parameters to be adjusted in order to fit the data points. Table \ref{tab-par-probegdist} shows these parameters for various DCM2 concentrations.

\begin{table}[htbp]
\centering
\caption{Optimized parameters used in eq. \ref{eq-probdcm2gap} to fit the data points show in Fig.\ref{fig-gap-distribution}.}
\begin{tabular}{ccccc}
\hline
\hline
DCM2               & $E_{0}$   & $\epsilon_{1}$ & $\epsilon_{2}$ & $\sigma$ \\
concentration (\%) & (eV) & (eV)     & (eV)     & (eV)   \\
\hline
1                  &  2.95944    &  0.10934   &  0.03800   &  0.02318   \\
2                  &  2.95904    &  0.11816   &  0.04500   &  0.02733   \\
5                  &  2.95605    &  0.12564   &  0.04900   &  0.02800   \\
10                 &  2.95735    &  0.14416   &  0.05100   &  0.03200   \\
\hline
\hline
\end{tabular}
\label{tab-par-probegdist}
\end{table}

Fig. \ref{fig-gap-distribution} shows that, surprisingly, for low DCM2 concentration the gap distribution does not depend significantly on the DCM2 concentration. Therefore, we conclude that the gap distribution is mostly determined by the random electric field produced by Alq$_{3}$ dipoles, differently from the usual understanding. Although Alq${_3}$ molecules have a smaller dipole moment, they are found more frequently near a given DCM2, thus explaining this behavior.

However, if this is the case, how can we understand the redshift due to increasing the DCM2 concentration? In Section \ref{kmc}, we present kinetic Monte-Carlo simulations of exciton dynamics\cite{fennel2012forster,madigan2006modeling} performed with the purpose of answering this question.

\section{Kinetic Monte-Carlo}
\label{kmc}

We propose that the emission redshift in Alq$_3$:DCM2 upon increasing DCM2 concentration is caused by diffusion and partial thermalization of excitons (limited by exciton lifetime\cite{herz2004time}). We propose that exciton diffusion in our system is described by FRET, which is a non-radiative energy transfer mechanism based on dipole-dipole coupling, where a donor molecule in an electronically excited state transfers its excitation energy to a nearby acceptor molecule \cite{lunz2011}. For efficient energy transfer, it is necessary that the emission spectrum of donor molecules overlaps the absorption spectrum of the acceptor molecules, and the separation distance between the donor and acceptor centers have to be much less than the wavelength \cite{Ren2016}. 

\begin{figure}[htbp]
\centerline{\includegraphics[width=8.0 cm]{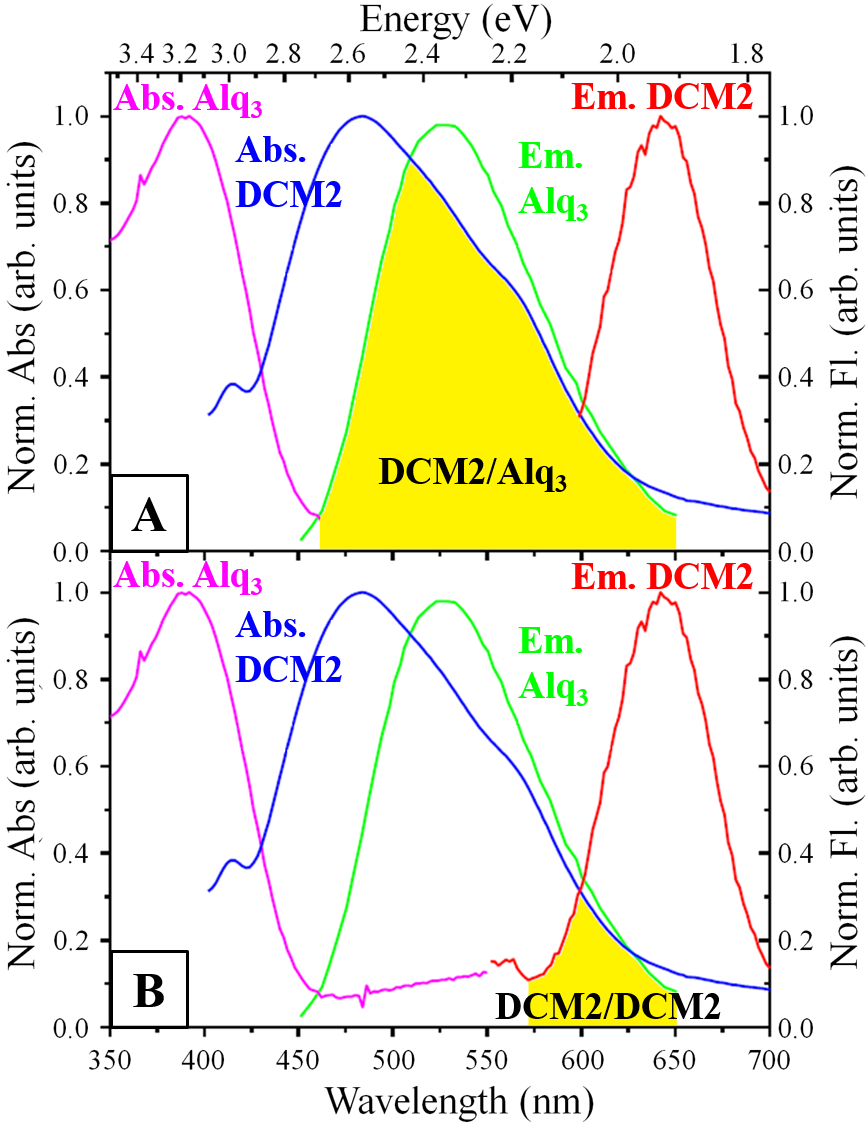}}
\caption{Experimental absorption and photoluminescence spectra for Alq$_{3}$ and DCM2 in a thin film of 50 nm thickness, obtained in this work. (a) Overlap (highlighted yellow region) between the Alq$_3$ emission and DCM2 absorption spectra, and (b) Overlap (highlighted yellow region) between the emission and absorption spectra of DCM2. These overlap areas (highlighted in yellow in both figures) are associated with the  F\"{o}rster radius in the FRET process.} 
\label{fig-forster-transfer}
\end{figure}

In our model, the exciton dynamics occur through two steps:

\begin{enumerate}
\item After exciton formation on a Alq$_{3}$ molecule (either by electric or photo-excitation), the excitation is quickly transferred to the nearest DCM2 molecule. This non-radiative energy transfer by F\"{o}rster mechanism is very efficient due to good spectral overlap between the donor (Alq$_{3}$) emission and acceptor (DCM2) absorption spectra, shown by the yellow region in Fig. \ref{fig-forster-transfer} (a). 

\item When excitons reach DCM2 molecules, or if they are initially formed directly on DCM2 molecules due to charge trapping, they can thermalize by hopping between DCM2 molecules also via F\"{o}rster process, since there is a smaller but non-negligible overlap between DCM2 emission and absorption spectra (Fig. \ref{fig-forster-transfer} (b)). Under energetic disorder, excitons move preferentially to lower energy sites. The thermalization process lasts until they finally decay radiatively (\textit{i.e}, after the exciton lifetime is reached, in average).       
\end{enumerate}

As stated above, the magnitude of spectral overlap between emission and absorption of donor and acceptor molecules is a key ingredient of the F\"{o}rster mechanism. We measure these quantities and the results are displayed in Fig. \ref{fig-forster-transfer}, which shows the experimental data for absorption and photoluminescence of an Alq$_{3}$:DCM2 matrix with concentration of guest material (DCM2) of 5\% into of host material (Alq$_{3}$). Both molecules were purchased from Lumtec (Luminescence Technology Corporation) and used without additional purification. The organic film was deposited in high vacuum environment (10$^{-6}$ Torr) by thermal evaporation onto quartz substrate and with a thickness of 50 nm. The quartz substrates were cleaned by ultrasonification using a detergent solution followed by ultrasonification with deionized water, followed by pure acetone, then pure isopropyl alcohol. For the organic layers the deposition rate was 0.5 \AA/s. UV-visible absorption spectra of the thin films were recorded using a Perkin-Elmer Lambda 950 dual-beam spectrometer with spectral correction. Thin film photoluminescence spectra were measured using a PTI fluorimeter model QuantaMaster 40 at room temperature and pressure conditions. The results  in Fig. \ref{fig-forster-transfer} show clearly the larger overlap for Alq$_3$-DCM2 with respect to DCM2-DCM2, thus justifying the larger F\"{o}rster radius used in simulations (see below) for the first case.

Due to the stochastic nature of the exciton hopping process, the exciton diffusion process is modeled by a kinetic Monte-Carlo method (kMC) based on the F\"{o}rster energy transfer (FRET), within the first-reaction method (FRM) approximation.\cite{madigan2006modeling,feron+12jap} The sample is modeled as a cubic lattice of 100$\times$100$\times$100 sites, with a certain proportion of DCM2 and Alq$_3$ sites given by the dopant concentration. The lattice constant is set to 1 nm. Then, $10^4$ excitons are randomly distributed in the cubic lattice and exciton dynamics simulation using the FRET process starts.

In the FRET model, the hopping time $t_{FRET}$ between any two sites $i$ and $j$ is given by:
\begin{equation}
t^{ij}_{FRET}=t_0\left( \frac{R_{ij}}{R_{0}}\right)^6\frac{1}{f(E_{i},E_{j})}
\label{eq-tfret}
\end{equation} 
\noindent where $t_0$ is the exciton lifetime, $R_{0}$ the F\"{o}rster radius, and $f(E_{i},E_{j})$ is a function accounting for energetic disorder. The exciton lifetime $t_0$ is 1.0 ns. The F\"{o}rster radius is proportional on the overlap integral of the donor emission spectrum (Alq$_{3}$) with the acceptor absorption spectrum (DCM2) (see Fig \ref{fig-forster-transfer} (a)). As there is a smaller overlap between DCM2 emission and absorption spectra (see Fig \ref{fig-forster-transfer} (b)), in the simulations we use two distinct $R_{0}$: One to account the jumps between Alq$_{3}$ and DCM2 ($R_0$=39 \AA), and another between DCM2-DCM2 molecules ($R_{0}$=6 \AA). The $R_{0}$ value for Alq$_3$/DCM2 energy transfer was taken from the literature \cite{deshpande1999white}, whereas the DCM2/DCM2 value was calculated using the ratio between the two yellow areas in Fig. \ref{fig-forster-transfer}.

The function $f(E_i,E_j)$ introduces the preferential hopping of excitons to lower energy sites and accounts for energetic disorder:
\begin{equation}
f(E_i,E_j) = \begin{cases}
\exp[\frac{-(E_j-E_i)}{k_BT}] &\text{if } E_j > E_i\\
1 &\text{if } E_j \leq E_i 
\end{cases}
\label{eq-energydis}
\end{equation}
The energies $E_i$ of all Alq$_{3}$ sites are randomly assigned according to a Gaussian distribution with a standard deviation, $\sigma$, extracted from Gaussians fitted to the Alq$_{3}$ absorption spectrum as described in Scheidler \textit{et al}. \cite{scheider+96prb} For DCM2 sites, the energies $E_i$ are randomly assigned according to the gap probability distribution function from Eq. \ref{eq-probdcm2gap}, obtained in Section \ref{efdist}.

\begin{figure}[htbp]
\centerline{\includegraphics[width=8.7 cm]{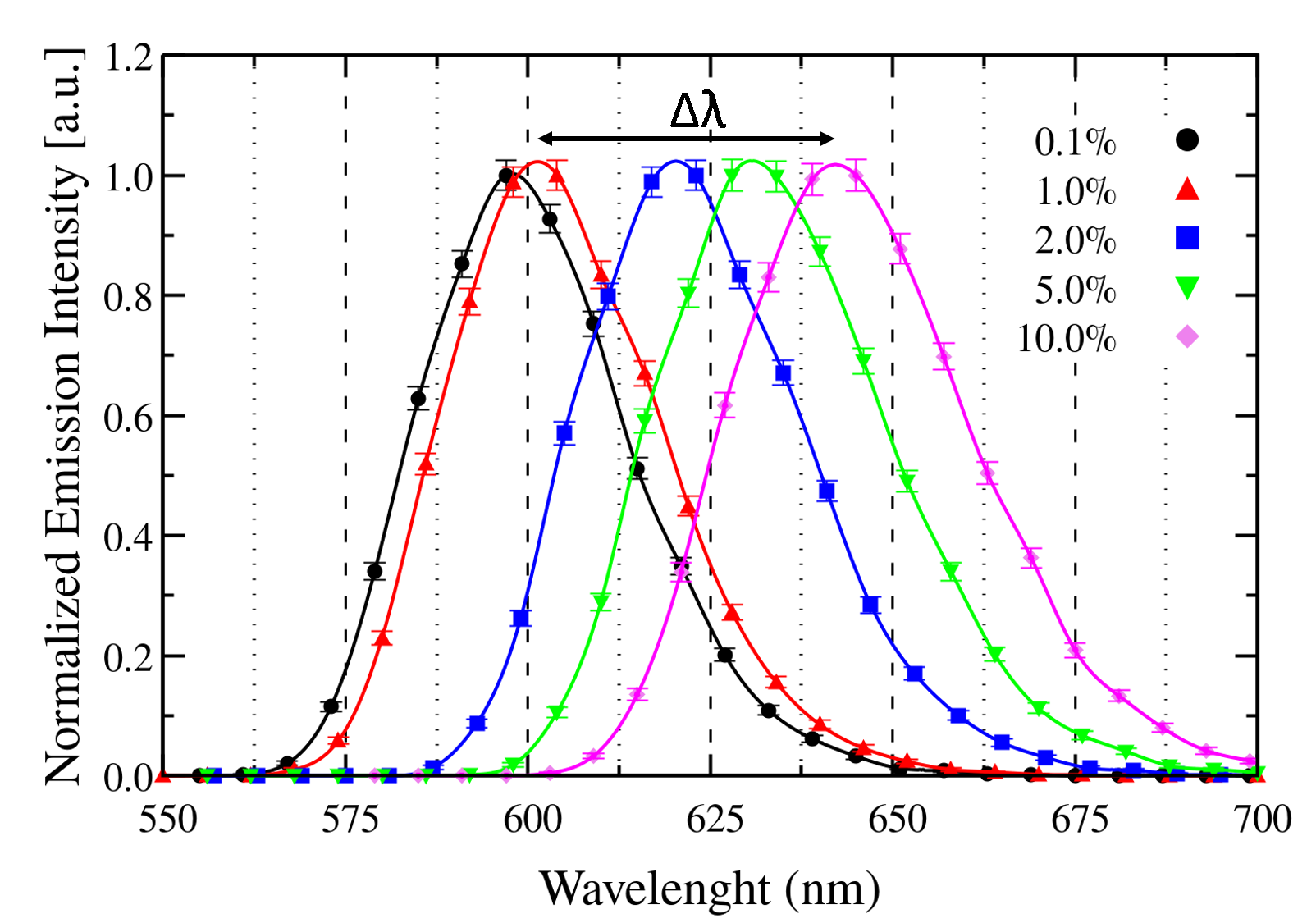}}
\caption{Normalized emission spectra obtained from KMC simulations for various DCM2 concentrations. The plot shows a redshift from 600 nm to around 645 nm, $\Delta\lambda$ $\sim$ 45 nm.} 
\label{fig-kmc-redshift-emission}
\end{figure}

In the first reaction method (FRM), a random number $X$ between 0 and 1 is selected for each process and a "jump time" is calculated:
\begin{equation}
t^{ij}_{jump}=-t^{ij}_{FRET}\ln X
\label{eq-jump-time}
\end{equation}
 The process with lowest jump time is then selected to be next destination of the exciton. The jump times for each exciton are summed and this process happens until the total event time reaches the exciton lifetime of 1 ns. When this occurs, we assume that the exciton is annihilated by emitting a photon. Then, the gap energy at the emission site is collected in a histogram (see Fig.\ref{fig-kmc-redshift-emission}). For all DCM2 molecules, the HOMO-LUMO gap energy at zero field $E_0$ is empirically redshifted by 0.25 eV to reproduce emission energy of the Alq$_3$:DCM2 system at very low DCM2 concentrations. In order to ensure the homogeneity of the DCM2 distribution, and to reduce the effects of the initial location of excitons, a total of 100 independent simulations were carried out for each concentration. Then, the final emission spectra is obtained as the average of all spectra obtained for a given concentration of DCM2.

All this theoretical effort culminates in the emission spectra shown in Fig. \ref{fig-kmc-redshift-emission}, as a function of DCM2 concentration. As the DCM2 concentration increases, the redshift in the emission spectra is observed. Experimental $\Delta\lambda$ $\sim$ 50 nm shift from 1\% to 10\% DCM2 concentration is reproduced \cite{bulovic+98cpl,madi+03prl}. In addition, the band width remains practically unchanged, as in experiments.

This is a very interesting result, since no assumption of local aggregation of DCM2 was needed and, as shown in the previous Section, the DCM2 gap distribution does not change considerably with concentration (in this low-concentration regime). Physically, we can understand the emission redshift as a consequence of the higher mobility of excitons when the DCM2 concentration increases: Within the exciton lifetime, for higher DCM2 concentrations, exciton DCM2-DCM2 jumps occur more frequently and therefore excitons have a better chance to thermalize to molecules with smaller gaps, thus causing an overall redshift of the average emission frequency. 

\section{Conclusions}
\label{conc}

In conclusion, using a combination of different theoretical methods and techniques, we propose a novel mechanism for the concentration-dependent emission redshift Alq$_{3}$:DCM2, based on exciton dynamics. Our theoretical modeling was composed of several important ingredients, that we now summarize: (1) DCM2 molecules suffer a nonlinear Stark shift of the electronic gap upon external electric fields, with a negative curvature (tendency to smaller gaps); (2) when DCM2 molecules are placed in an Alq$_3$ matrix, the random dipole moments of neighboring molecules produce local electric fields that generate a distribution probability of DCM2 with a long tail towards low energies. For low DCM2 concentrations, these local fields are caused primarily by Alq$_3$ molecules, differently from the usual understanding. (3) Exciton hopping from Alq$_3$ to DCM2 and specially between DCM2 molecules allow thermalization of excitations towards lower energies and explain the redshift. For larger concentrations of DMC2, exciton mobility is larger and therefore the redshift is more substantial. Our model agrees quantitatively with experiments and we believe it describes a very general mechanism that should occur in similar systems.  

\section{Acknowledgements}

The authors acknowledge financial support from Brazilian agencies CNPq, FAPERJ, Finep, INCT - Nanomateriais de Carbono and INCT-INEO for financial support. Ronaldo Giro wish to thank Dr. Ulisses Mello, director of IBM Research - Brazil, for partial support of this project and to his colleagues in the Smarter Devices team for many stimulating discussions. Grazi\^ani Candiotto gratefully acknowledge FAPERJ Processo E-26/200.008/2020 for financial support. The authors wish to thanks Dr. Juan H. S. Restrepo from Universidad Pontificia Bolivariana, Medellin, Colombia for providing the experimental data. The authors also acknowledge the support of N\'ucleo Avan\c{c}ado de Computa\c{c}\~ao de Alto Desempenho (NACAD/COPPE/UFRJ), and Sistema Nacional de Processamento de Alto Desempenho (SINAPAD).

\bibliographystyle{prsty}
\bibliography{dcm2-bib}

\end{document}